\documentclass[journal,twoside]{IEEEtran}
\usepackage{hyperref}
\usepackage{amsmath}
\usepackage{graphicx}
\usepackage{multicol}
\usepackage{balance}
\usepackage{cite}
\usepackage{flushend}
\usepackage{subfigure}
\usepackage{color}
\usepackage{comment}

\begin{document}
	
\title{Optimal Coupling Patterns in Interconnected Communication Networks
\thanks{Manuscript received December 10, 2017. This work was supported by the National Natural Science Foundation of China under Grant No. 61503420.}
	}
\author{{Jiajing Wu, {\em Member, IEEE,} Jian Zhong, and Zhenhao Chen}
\thanks{The authors are with the School of Data and Computer Science, Sun Yat-sen University, Guangzhou 510006, China. (Email:  wujiajing@mail.sysu.edu.cn)} 

\thanks{Digital Object Identifier 00.0000/TCSII.2017.000000}
	}	
\markboth{IEEE TRANSACTIONS ON CIRCUITS AND SYSTEMS--II: EXPRESS BRIEFS,~VOL.~XX, No.~XX,~xx~2017}%
{Wu \MakeLowercase{\textit{et al.}}: Optimal Coupling Patterns in Interconnected Communication Networks}
	\maketitle
	
\begin{abstract}
Traffic dynamics on single or isolated complex networks has been extensively studied in the past decade. Recently, several coupled network models have been developed to describe the interactions between real-world networked systems. In interconnected communication networks, the coupling links refer to the physical links between networks and provide paths for traffic transmission. 
In this paper, we employ a simulated annealing (SA) algorithm to find a near-optimal configuration of the coupling links, which effectively improves the overall traffic capacity compared with random, assortative and disassortative couplings. Furthermore, we investigate the optimal configuration of coupling links given by the SA algorithm and develop a faster method to select the coupling links.

\end{abstract}

\begin{IEEEkeywords}
Complex networks, interconnected networks, coupling pattern, traffic transmission.                   
\end{IEEEkeywords}

\section{Introduction\label{sec:Intro}}
\IEEEPARstart{N}{owadays}, the efficient operation of critical networked infrastructures, such as the Internet, power grids, transportation networks and communication networks, has become vital for maintaining all essential activities of society~\cite{Wang2003}. Because of the ever-increasing demand of networked resources, many networked infrastructures would exhibit a phase transition point where the network transits from a free-flow state to a congestion state~\cite{Sreenivasan2007}. 

It is universally acknowledged that the random graph model proposed by Erd\"os and R\' enyi~\cite{Erdos1959} in the late 1950s opened up the systematic study of complex network structure in mathematics. Since the discovery of small world~\cite{Watts1998} and scale-free~\cite{Barabasi1999} topological features at the end of the last century, there have been a great deal of studies on understanding the interplay between network structure and dynamics from the perspective of complex networks~\cite{Xia2008,Zhang2014,Liu2015}. 

Although the problem of traffic congestion has been extensively studied, the focus of most previous work has been on isolated networks. However, many infrastructure networks in the real world are actually coupled together or interacting with each other. In order to model the interactions between real-world networks, several coupled network models have been developed~\cite{Xia2016,Sole-Ribalta2016,Chen2017}. In these network models, events taking place in one system are likely to have impacts on others.

For example, a theoretical model of interdependent networks was proposed in~\cite{Buldyrev2010} to model the interdependency between two networks. In this work, interdependency links between two random graphs were used to represent the logical interdependency between two networks. 
As another kind of coupling model, the interconnected networks contains coupling links that are physical links between networks and provide paths for traffic transmission~\cite{interconnected,Xia2016}. 

Inspired by previous studies of traffic congestion in isolated complex networks and the newly developed concepts of interconnected networks, in this paper, we explore the optimal coupling pattern to minimize traffic congestion in interconnected complex networks based on a data-packet transport model.
Previous studies~\cite{Wu2013,Wu2015} have demonstrated that, to alleviate traffic congestion and improve the overall network performance, the traffic loads should be uniformly distributed in the network and the average data transmission distance should be short. 
Based on this criterion, we employ the simulated annealing algorithm to find the near-optimal way to place the coupling links subject to the maximal network transmission capacity. 


\section{Network Model\label{sec:model}}
In this paper, we consider packets being sent in discrete time steps. At each time step, $\lambda N$ new packets are generated with randomly sources and destinations, where $\lambda$ is the packet generation rate of each node and $N$ is the number of nodes in the network. At the same time, each node receives packets from its adjacent nodes. Each node has a buffer queue that stores the packets waiting to be processed and all the incremental packets will be pushed into the buffer queue according to the First-In-First-Out principle. Each packet with a particular pair of source and destination is always transmitted along a pre-calculated shortest routing path. Denoting the transmission capacity of each node as $R$, at each time step, the first $R$ packets of each node are transmitted by one step to their destinations according to their routing paths. 
Packets that have already reached their destinations are removed from the network.

It has been demonstrated that~\cite{Sreenivasan2007}, as the traffic intensity increases, a phase transition occurs, taking the network from a free-flow state to a congestion state. We define the \textit{critical generation rate} $\lambda_c$ as an indicator of the network transmission capacity, which equals the average number of the newly generated packets per node per time step when the phase transition occurs in the network. A larger $\lambda_c$ implies that the network can handle higher traffic intensity without congestion.\par

In our previous work~\cite{Wu2013}, we have given a theoretical
estimation of the critical generation rate $\lambda_c$ in a single network, which can be expressed as
\begin{equation}
\lambda_c=\frac{R}{\tilde{D}U(i)_{\textrm{max}}N},
\label{eqn:lambda_c}
\end{equation}
where $R$ is node transmission capacity, $\tilde{D}$ is the average transmission distance of each packet, and $U(i)$ is the node usage probability of node $i$, which is defined as
\begin{equation}
U(i)=\frac{\sum\limits_{u,w\in{V},\atop u\neq{w}\neq{i}}\sigma_{uw}(i) }{\sum\limits_{j\in{V}}\sum\limits_{u,w\in{V},\atop u\neq{w}\neq{j}}\sigma_{uw}(j)},
\label{eqn:usagepro}
\end{equation}
where $V$ is the set of all nodes in the network, $\sigma_{uw}(i)$ is defined as 1 if node $i$ lies on the routing path between node $u$ and $w$, and as 0 otherwise.

The total number of paths that pass through node $i$, denoted by $C(i)$, can be expressed as
\begin{equation}
C(i)=\sum\limits_{u,w\in{V},\atop u\neq{w}\neq{i}}\sigma_{uw}(i),
\label{eqn:ci}
\end{equation}

And the average transmission distance $\tilde{D}$ can be approximated as
\begin{equation}
\tilde{D}\approx\frac{\sum\limits_{j\in{V}}C(j)}{N(N-1)},\label{eqn:avrdist}
\end{equation}
where $N$ is the total number of nodes in the network.

From (\ref{eqn:usagepro}), (\ref{eqn:ci}) and \ref{eqn:avrdist}), we have
\begin{equation}
U(i)\tilde{D}\approx\frac{C(i)}{N(N-1)},\label{eqn:multiply}
\end{equation}

Substituting it into (\ref{eqn:lambda_c}), we have
\begin{equation}
\lambda_c\propto\frac{1}{C_{\textrm{max}}},\label{eqn:cmax}
\end{equation}
where $C_{\textrm{max}}$ is the maximum value of $C(i)$ in the network.\par

Therefore, a larger $C_{\textrm{max}}$, which is defined as the maximum value of $C(i)$, implies a larger $\tilde{D}B_{\textrm{max}}$ and a smaller $\lambda_c$.

\section{Coupling Patterns \label{sec:prefer}}
Without loss of generality, here we consider two networks, labelled $A$ and $B$. Note that both networks are packet transmission networks which work under the operation model described in Section~\ref{sec:model}. For simplicity and clarity of the results, we assume that these two networks are of the same size $N = N_A= N_B$ and the same average degree $\langle k \rangle$ = $\langle k_A \rangle$ = $\langle k_B \rangle$. 
The coupling probability $P$ is defined as the ratio between the number of interconnected links and the network size. Besides the density of interconnected links, it has been widely demonstrated~\cite{Tan2014,Du2015dynamic} that coupling patterns, i.e., the ways interconnected links are added, also have significant influence on the dynamical processes of interconnected networks. In much previous work, three kinds of coupling patterns based on the heterogeneity of traffic loads have been considered~\cite{Tan2013}.

\begin{enumerate}
	\item {Assortative coupling (AS)}. We sort nodes in networks $A$ and $B$, both in descending order of node usage probability. 
We connect the first node in network $A$ with the first node in network $B$, and then connect the second node in network $A$ with the second node in network $B$, and so on. Repeat this process until all coupling links are built.
	\item {Disassortative coupling (DIS).} We sort nodes in network $A$ ($B$) in descending (ascending) order of node usage probability. 
We connect the first node in network $A$ with the first node in network $B$, and then connect the second node in network $A$ with the second node in network $B$, and so on. Repeat this process until all coupling links are built.
	\item {Random coupling (RD).} Randomly choose a node in network $A$ and a node in network $B$. If neither has a coupling link, then connect them. Otherwise randomly choose another two coupling links and check whether repeated links exist. Repeat this process until all coupling links are built.
\end{enumerate}\par

As demonstrated in the previous section, a smaller $C_{\textrm{max}}$ corresponds to a larger $\lambda_c$. This analytical result remains universal for 
interconnected networks because the interconnected links provide routing paths for traffic transmission between two networks.
Therefore, to mitigate traffic congestion and achieve maximum transmission capacity, we adopt a nature-inspired optimization method, namely, simulated annealing (SA)~\cite{Kirkpatrick1983}, to find the optimal coupling pattern to make $C_{\textrm{max}}$ of the interconnected networks as small as possible. The procedure of the algorithm is as follows:
\begin{enumerate}
\item Adopt the above-mentioned random coupling to add coupling links between the two networks. Calculate $C_{\textrm{max}}^0$ of the initial networks and set $C_{\textrm{max}}^{\textrm{best}}=C_{\textrm{max}}^0$. Set the time $t=0$ and the epoch count $g=0$.
\item Randomly select a coupling link $e_{ij}$, which connects node $i$ from network $A $ with node $j$ from network $B$, and remove it. Then randomly select a node $x$ from network $A$ and a node $y$ from network $B$. Connect node $x$ with node $y$ if there is no existing link between them, otherwise randomly choose another pair of nodes to connect. Note that the \textit{temperature} \textit{T} does not change in this step.
\item Recalculate $C_{\textrm{max}}$ after rewiring and denote it as $C_{\textrm{max}}^{\textrm{new}}$. If $C_{\textrm{max}}^{\textrm{new}}<C_{\textrm{max}}^{\textrm{best}}$, accept the new coupling link. If $C_{\textrm{max}}^{\textrm{new}}\geq C_{\textrm{max}}^{\textrm{best}}$, accept the link rewiring with the probability $e^{-{\Delta}/{T}}$, where \textit{T} is a control parameter called \textit{temperature} and $\Delta=C_{\textrm{max}}^{\textrm{new}}-C_{\textrm{max}}^{\textrm{best}}$. If the rewiring is accepted, set $C_{\textrm{max}}^{\textrm{best}}=C_{\textrm{max}}^{\textrm{max}}$, $t=0$ and $g=g+1$; otherwise, set $t=t+1$ and keep $g$ unchanged.
\item Change the temperature $T$ to $\alpha T$ every 20 epoch counts, where $\alpha \left(0 < \alpha < 1 \right)$ is a parameter called the cooling ratio.
\item Iteratively execute steps 2 to 4 until time $t$ is smaller than 10000, which implies that $C_{\textrm{max}}^{\textrm{best}}$ is unchanged in the latest 10000.
\end{enumerate}\par

In the initial stage, the temperature parameter $T$ should be large enough and the cooling ratio $\alpha$ should be as close to 1 as possible to allow the search of solutions to escape from local optima. In our simulations, we set the initial temperature $T_0=1500$ and $\alpha = 0.999$.

\section{Network Performance\label{sec:result}}
In this section, to evaluate the effects of coupling patterns on traffic congestion in interconnected networks, we explore two types of network models, namely coupled Barab\'asi-Albert scale-free networks (BA-BA) and coupled Erd\"os-R\'enyi networks (ER-ER). For each scenario, we perform simulations on two networks of equal size $N_A$ = $N_B$ = 500 and average degree $k_A$ = $k_B\approx$ 5.976. In our simulations, each node has the same transmission capacity per time step $R = 1000$, and the coupling probability $P$ varies from 0.1 to 1.0.

\subsection{Performance Indicators}
In this paper, we consider three performance indicators, namely, critical generation rate $\lambda_c$, average transmission distance $\tilde{D}$, and closeness of interconnected networks $C$. 

As mentioned in Section~\ref{sec:model}, the critical generation rate $\lambda_c$ characterizes the phase transition point between the congestion state and the free-flow state, which can be viewed as an indicator of the network transmission capacity. The average transmission distance $\tilde{D}$ is defined as the average number of hops between each pair of nodes and can be used to characterize the transmission efficiency. Moreover, to indicate how closely the two networks are interconnected, we propose to use the closeness of coupled networks $C$, which is defined as
\begin{equation}
C=\sum\limits_{i\in{V_A}}\sum\limits_{u,w\in{V_B},\atop u\neq{w}}R_{\rm uw}(i)+\sum\limits_{j\in{V_B}}\sum\limits_{m,n\in{V_A},\atop u\neq{w}}R_{\rm mn}(j),
\label{eqn:closeness}
\end{equation}
where $V_A$ and $V_B$ are the sets of the nodes in networks $A$ and $B$, respectively, $R_{\rm uw}(i)$ is defined as 1 if node $i$ from network $A$ lies on the routing path between nodes $u$ and $w$ from network $B$, and as 0 otherwise. 
According to the definition given in Eq.~(7), we can see that $C$ quantifies the number of paths between any two nodes of one network passing a node of the other network. A large value of $C$ indicates that the internal transmission of one network is affected by the other network to a greater extent, and thus these two networks can be regarded as more intimately connected. 

\subsection{Simulation Results}

\begin{figure}[t]
	\centering
	\includegraphics[width=0.24\textwidth,height=1.6in, trim=0 0 0 30]{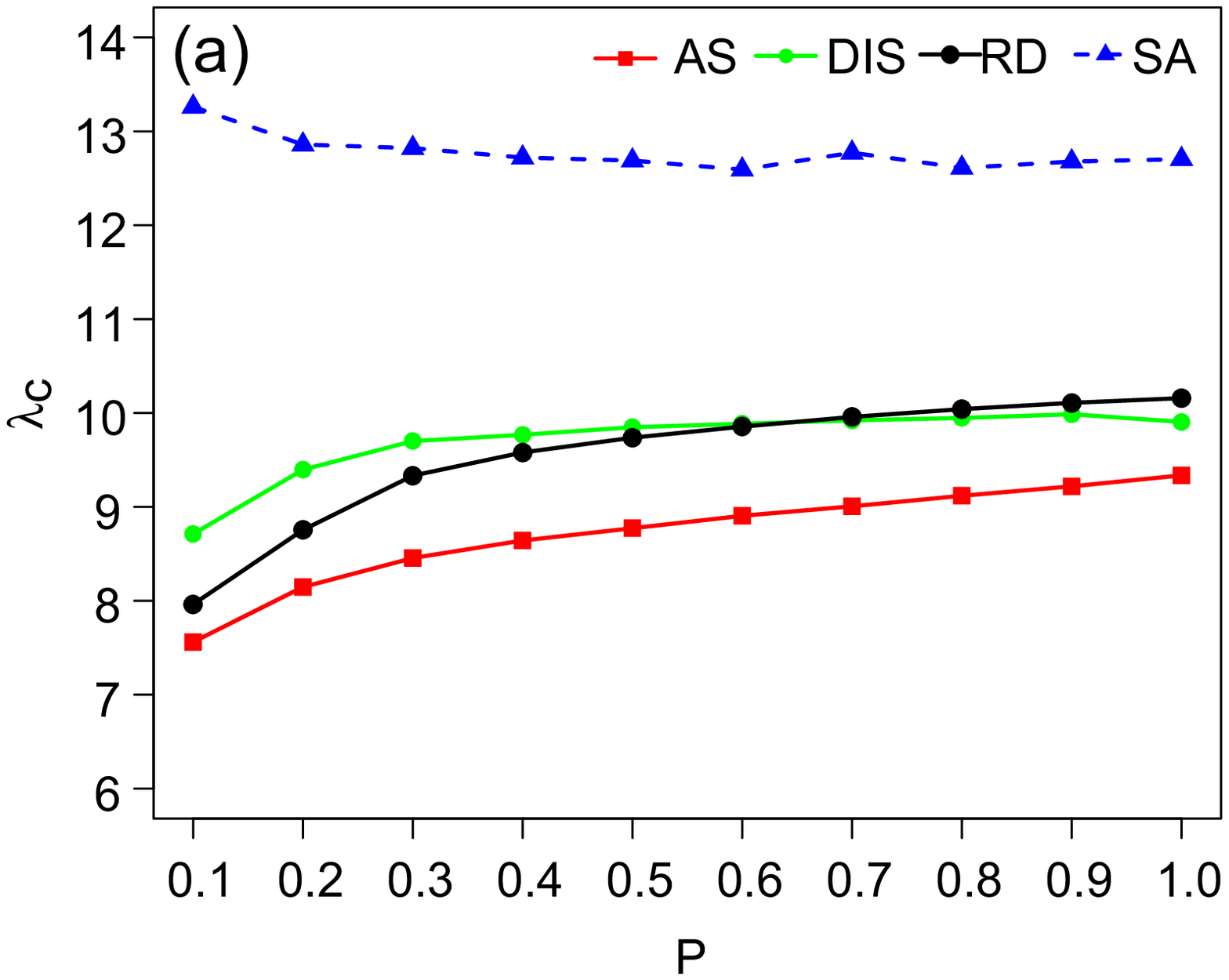}
		\vspace{-0.3cm}
	\includegraphics[width=0.24\textwidth,height=1.6in, trim=0 0 0 30]{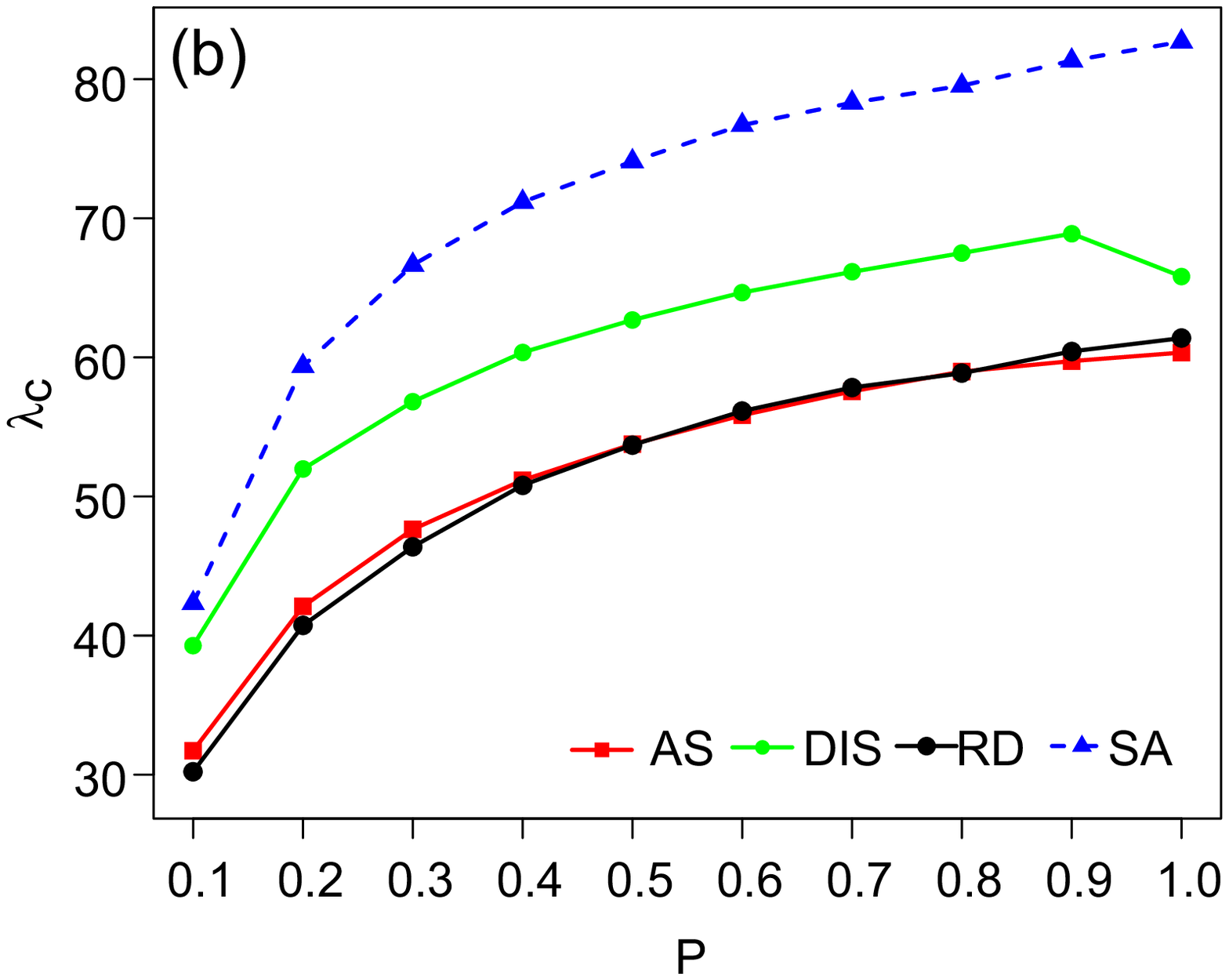}
	\caption{Critical generation rate $\lambda_c$ versus coupling probability $P$ under AS, DIS, RD and SA couplings, for BA-BA(a) and ER-ER(b) interconnected networks with network size $N_A=N_B=500$. Each point is averaged over 50 independent runs.}
	\label{Fig.2}
\end{figure}

Fig.~\ref{Fig.2} exhibits how the traffic capacity $\lambda_c$ evolves with different coupling patterns and probabilities, for BA-BA and ER-ER interconnected networks. We can observe that, for all cases, the SA algorithm performs much better than the other three couplings. Besides, $\lambda_c$ of the DIS coupling is larger than that of the AS coupling in general. The reason is that AS coupling which connected the high degree nodes from each side together will lead to more severe traffic concentration on hub nodes, especially when the coupling probability $P$ is relatively low. Contrarily, when the networks are DIS coupled, the nodes with low node degree (usually carry low traffic loads) can release part of traffic loads of the high-degree nodes and effectively balance the distribution of traffic loads.

\begin{figure}[t]
	\centering
	\includegraphics[width=0.24\textwidth,height=1.6in, trim=0 0 0 30]{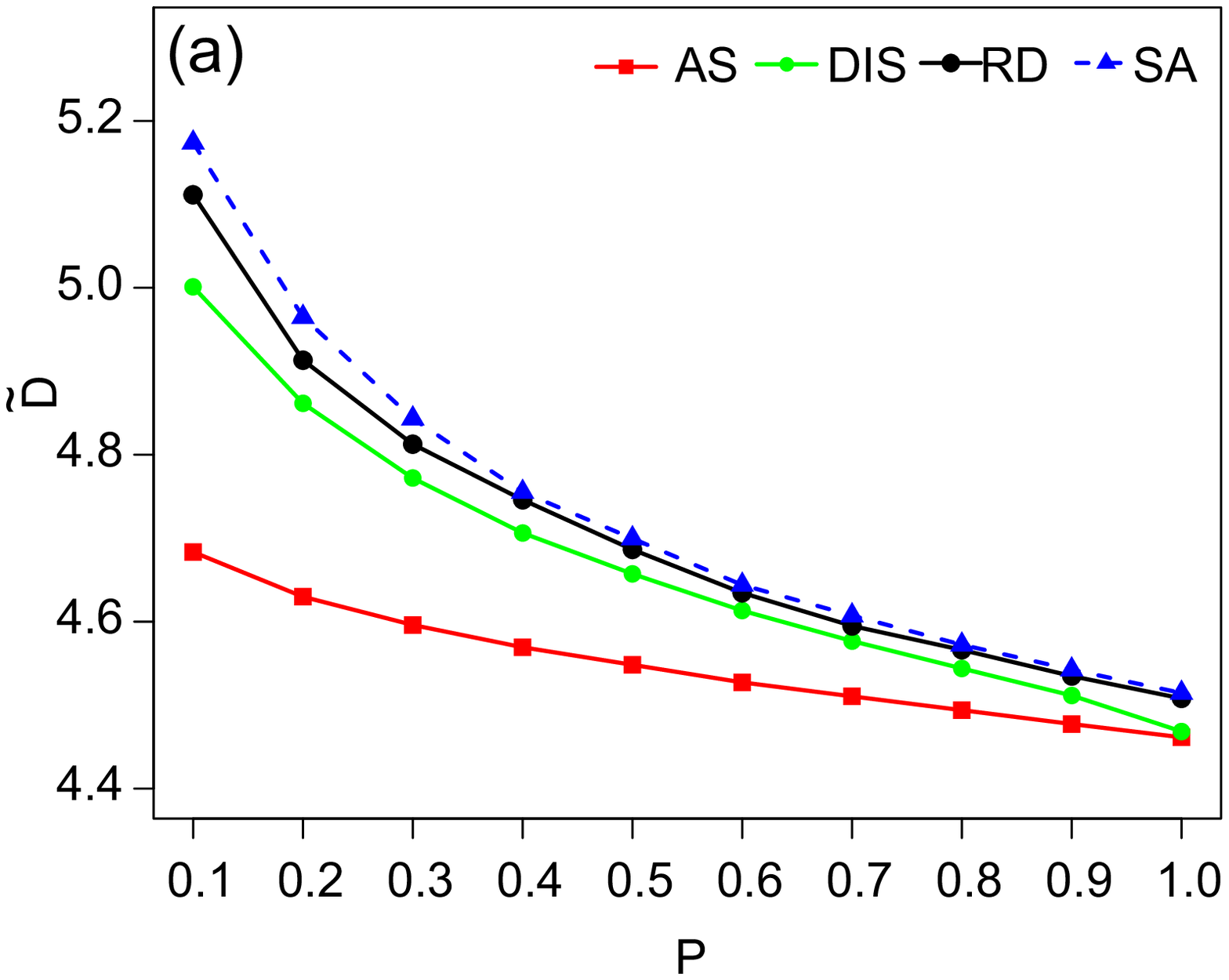}
	\vspace{-0.3cm}
	\includegraphics[width=0.24\textwidth,height=1.6in, trim=0 0 0 30]{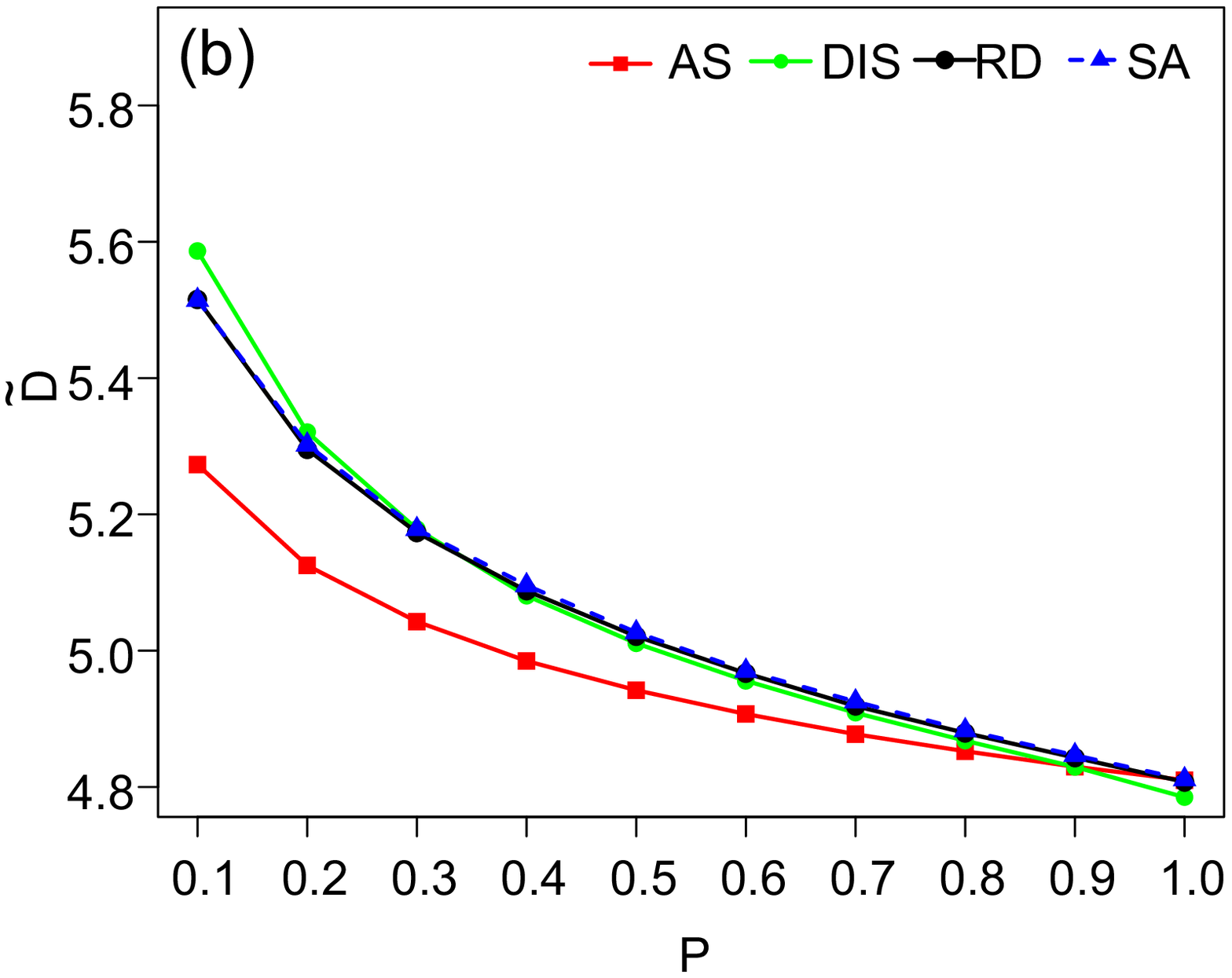}
	\caption{Average transmission distance $\tilde{D}$ versus coupling probability $P$ under AS, DIS, RD and SA couplings, for BA-BA(a) and ER-ER(b) interconnected networks with network size $N_A=N_B=500$. Each point is averaged over 50 independent runs.}
\label{Fig.3}
\end{figure}

\begin{figure}[b]
	\centering
	\includegraphics[width=0.24\textwidth,height=1.5in, trim=0 0 0 60]{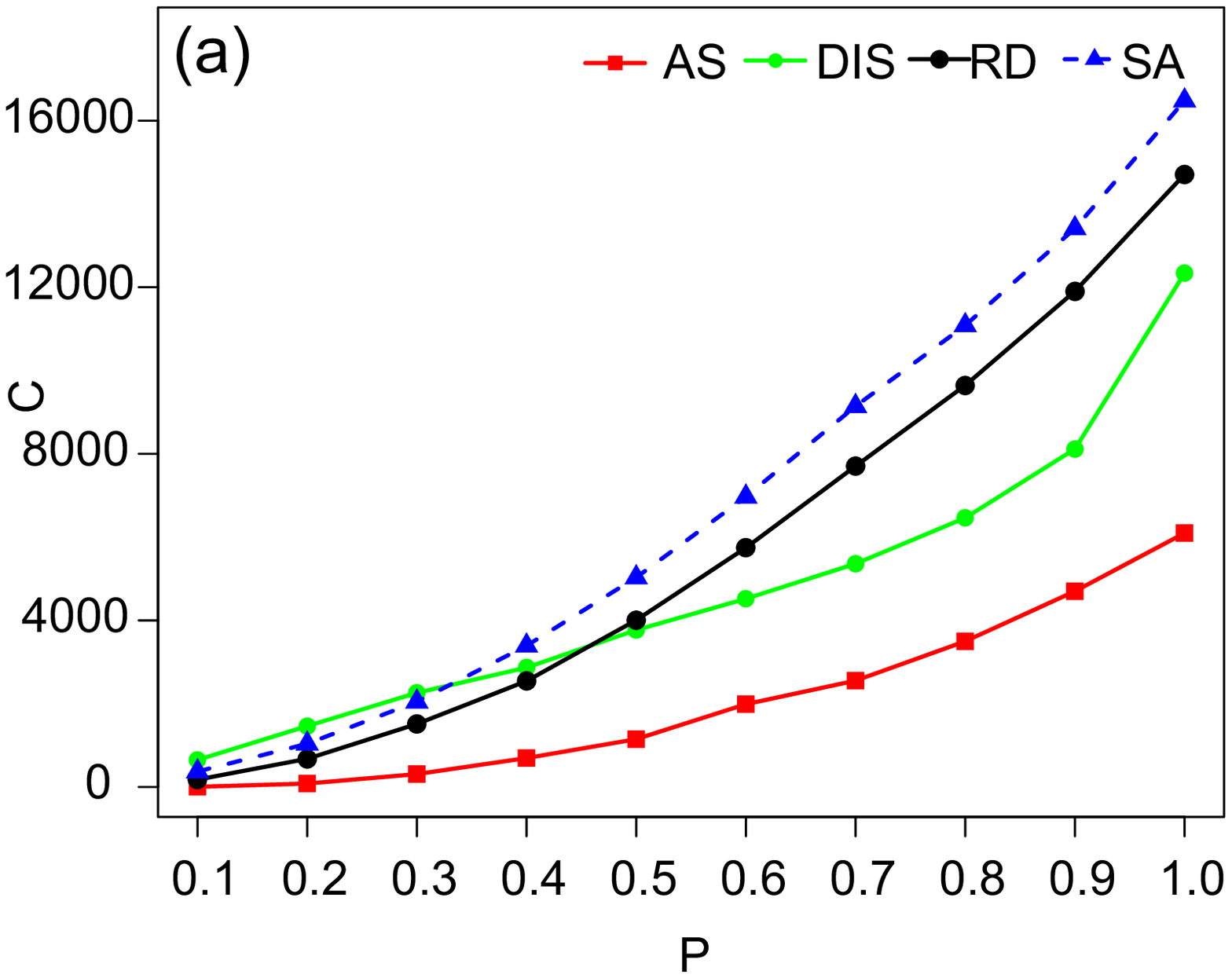}
	\vspace{-0.3cm}
	\includegraphics[width=0.24\textwidth,height=1.5in, trim=0 0 0 60]{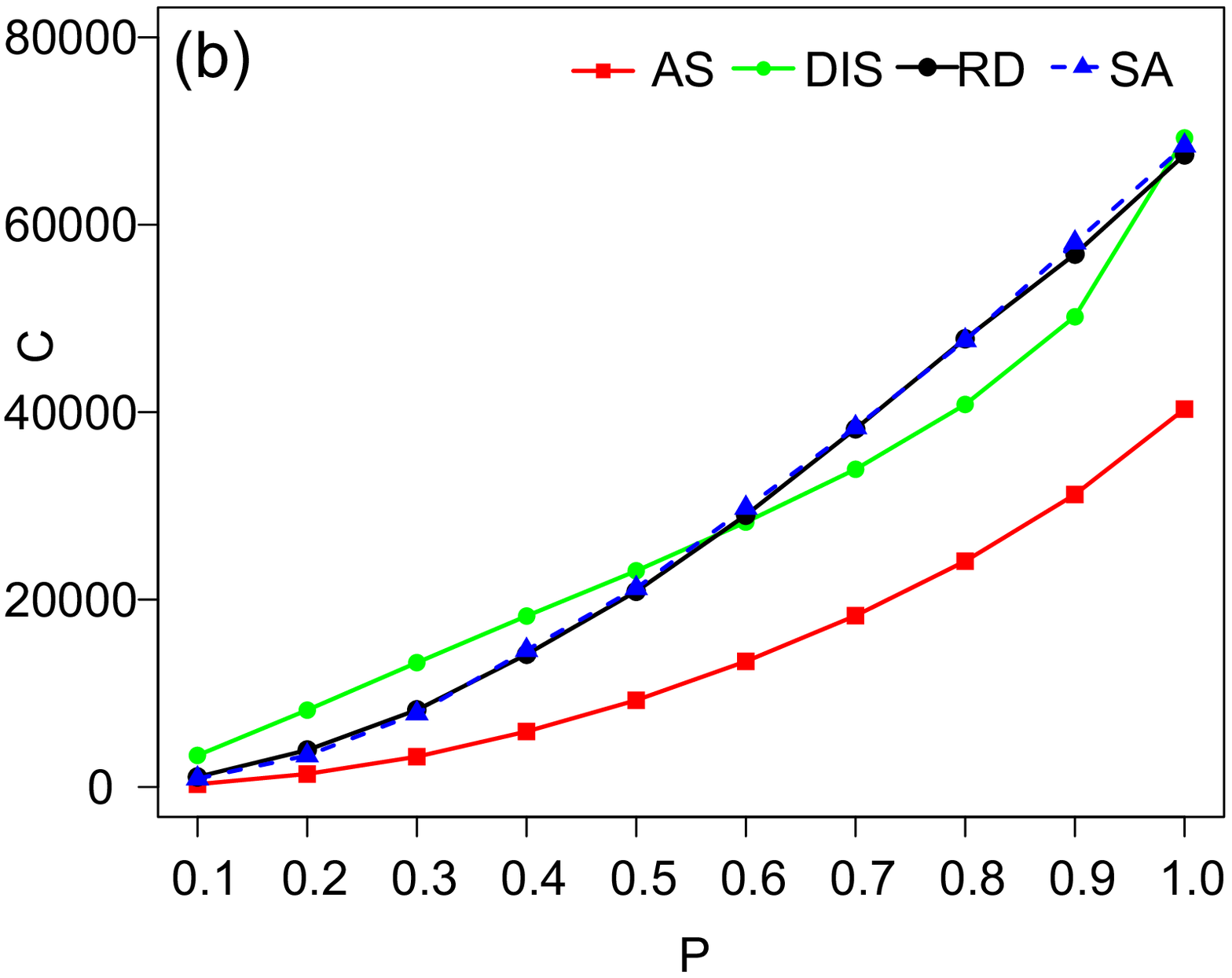}
    \caption{Closeness of interconnected networks $C$ versus usage probability $P$ under AS, DIS, RD and SA couplings, for BA-BA(a) and ER-ER(b) interconnected networks with network size $N_A=N_B=500$. Each point is averaged over 50 independent runs. }
	\label{Fig.4}
\end{figure}

As shown in Figs.~\ref{Fig.3} and \ref{Fig.4}, the average transmission distance $\tilde{D}$ declines and the closeness of coupled networks $C$ rises with the increase of the coupling probability $P$. In other words, more coupling links between two originally isolated networks can make them more closely connected. Moreover, we can observe that for each type of interconnected networks, AS coupling performs best in terms of the average transmission distance $\tilde{D}$ and worst in terms of the closeness $C$. If we adopt the SA algorithm to add the coupling links, the two networks are most closely connected with a largest value of $C$.

Next, we make comparisons between different network types. From Figs.~\ref{Fig.2},~\ref{Fig.3} and~\ref{Fig.4}, one can see that, with the same coupling pattern and probability, ER-ER networks outperforms BA-BA networks in terms of $\lambda_c$ and $C$, but perform worse in terms of $\tilde{D}$. That is to say, when the networks being connected are more homogeneous, it would be easier to achieve a large transmission capacity and make the networks more closely connected at the expense of a slightly longer transmission time.

To put our work in a more practical context, we also implement simulations on interconnected Internet Autonomous-System(AS)-level networks of South Korea (SK) and Japan (JP). We acquire the topological data from an on-line dataset provided by the Cooperative Association for Internet Data Analysis~(CAIDA)~\cite{Tan2014}~\cite{CAIDA}. 
Networks SK and JP are with network sizes $N_{SK}$ = 677 and $N_{JP}$ = 509. These two networks have rather different average internal degrees ($k_{SK}\approx$ 3.65 and $k_{JP}\approx$ 4.40), and they are sparsely interconnected by just 14 external coupling links. Table~\ref{table_1} compares the simulation results under AS, DIS, RD and SA couplings, as well as the realistic coupling information named CAIDA here. We can observe that, similar to the results for coupled BA scale-free networks when $P$ is closed to 0, AS coupling leads to the shortest $\tilde{D}$, DIS coupling gets the largest $C$, and SA performs best in terms of $\lambda_c$.

Therefore, we can conclude that there exists a trade-off: if we want to shorten the value of $\tilde{D}$, the coupling links should be added between high-degree nodes from both sides; on the contrary, if we aim to suppress traffic congestion and make the interconnected networks more integrated as a whole, we can adopt the DIS coupling or the SA algorithm to add coupling links between the networks.

\begin{table}[h]
\renewcommand{\arraystretch}{1.5}
\caption{Critical generation rate $\lambda_c$, average transmission distance $\tilde{D}$, and closeness $C$ under different coupling patterns for interconnected Internet AS-level networks of South Korea and Japan.}
\label{table_1}
\centering
\begin{tabular}{c|c c c c c }
\hline
 &AS &DIS &RD &CAIDA &SA  \\ 
\hline
\hline
$\lambda_c$ & 2.4627 & 2.6480  & 2.2013 & 2.3087 & 3.2847    \\
\hline
$\tilde{D} $ & 4.0860 & 4.6537  & 4.7386 & 4.3859 & 4.8606   \\
\hline
$C$ & 0 & 122 & 10 & 114  & 68     \\ 
\hline
\end{tabular}
\end{table}

\section{Heuristic Methods\label{sec:SP}}

As shown in the previous section, the proposed SA coupling algorithm can effectively enhance the traffic capacity. However, the convergence time of the SA algorithm increases with the network sizes and the number of coupling links, and thus this strategy is quite time consuming for large-scale networks due to its high complexity. Therefore, in this section, we examine the optimal configurations of the SA-selected coupling links and develop a faster method to select the coupling links. 



\begin{figure}[t]
	\centering
	\includegraphics[width=0.24\textwidth,height=1.6in, trim=0 0 0 30]{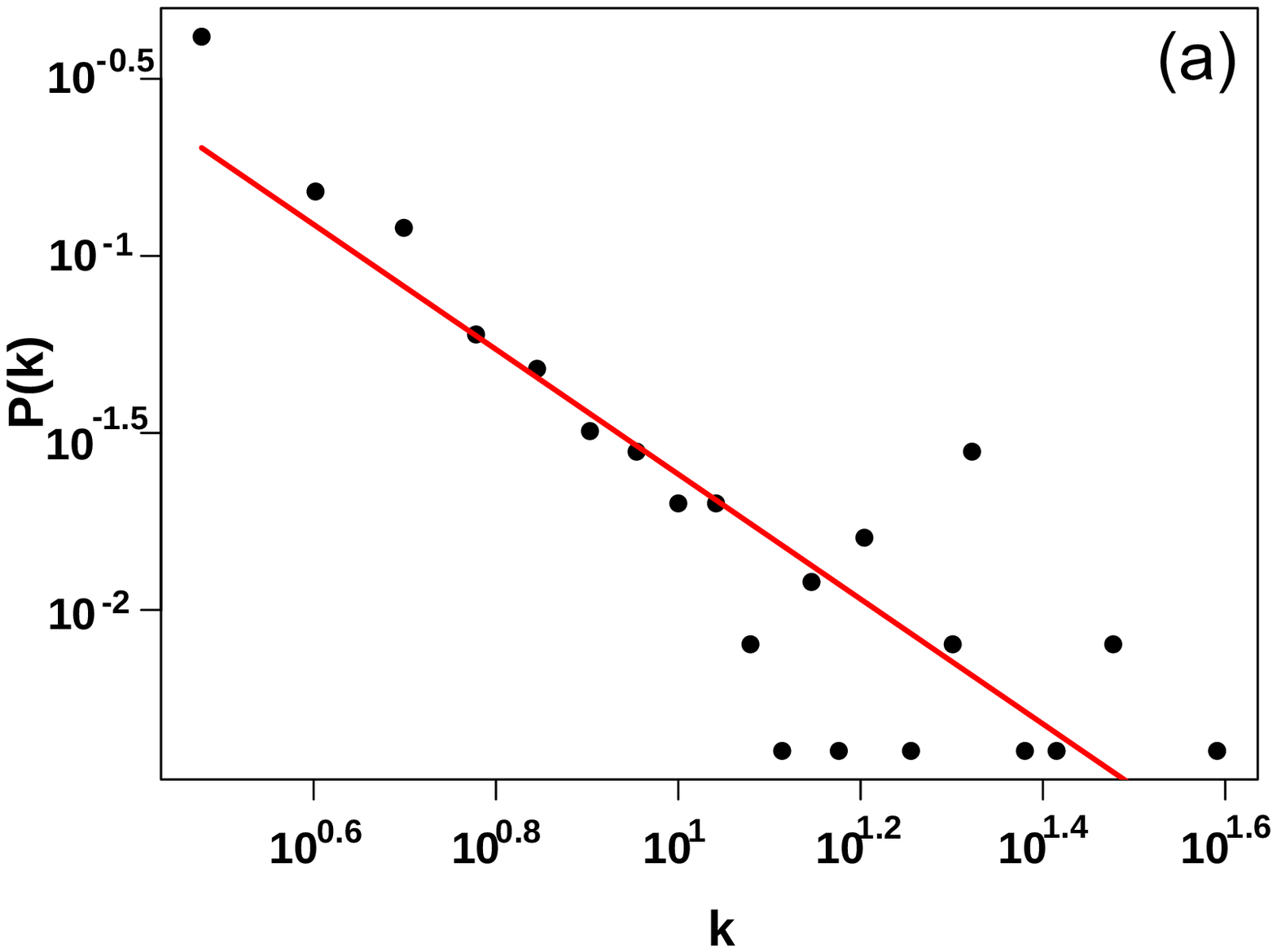}
		\vspace{-0.3cm}
	\includegraphics[width=0.24\textwidth,height=1.6in, trim=0 0 0 30]{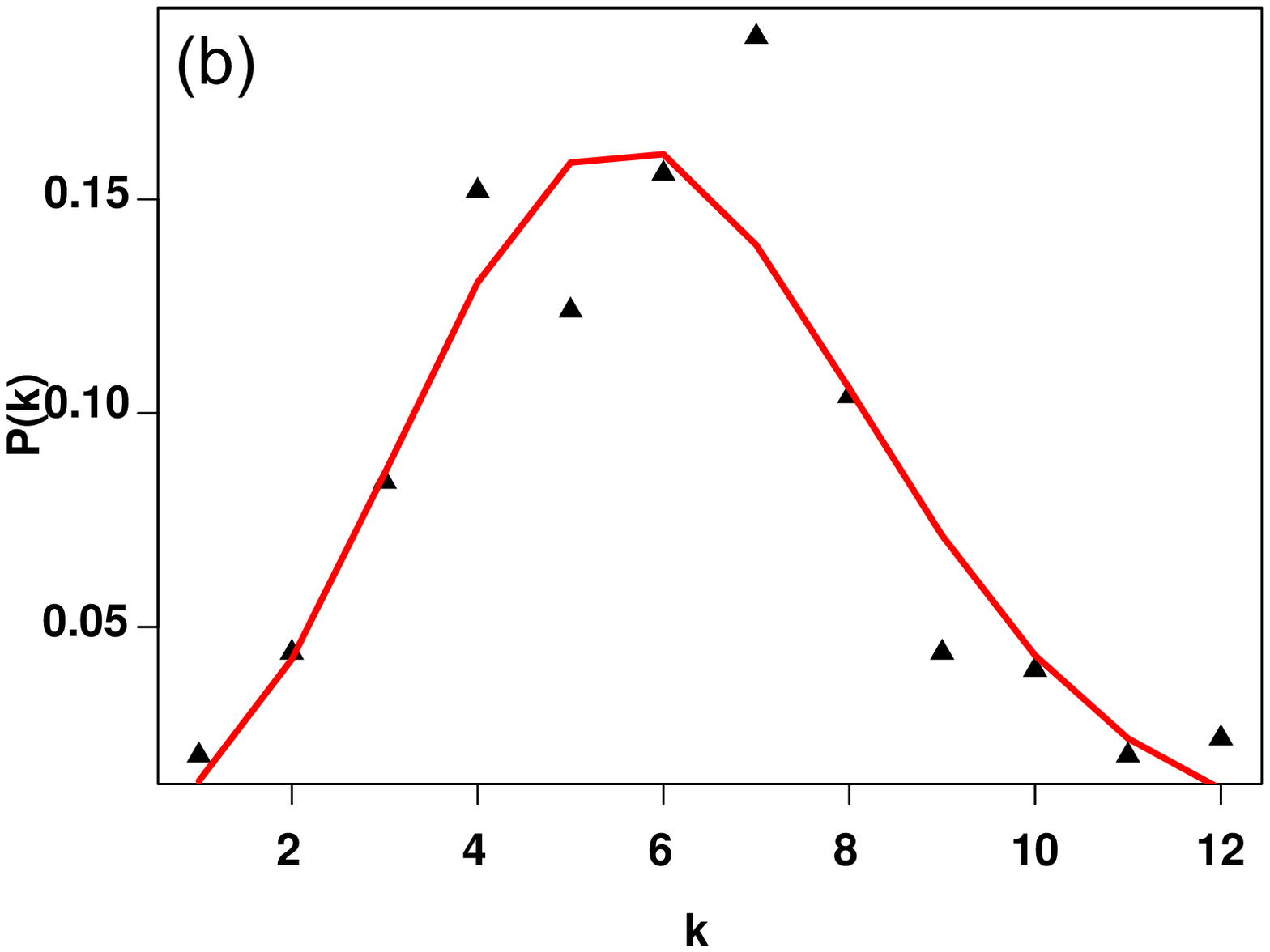}
	\caption{Fitting of the degree distribution of the nodes with coupling links for BA-BA(a) and ER-ER(b) interconnected networks with network size $N_A$ = $N_B$ = 500 and coupling probability $P$ = 0.5 on a log-log scale.}
	\label{Fig.6}
\end{figure}

Fig.~\ref{Fig.6}(a) shows the degree distribution of the coupling nodes in a BA scale-free network. For the BA scale-free networks, most coupling links are added between the nodes with relatively low degrees. We plot the degree distribution in a log-log scale, and its shape is similar with a power-law distribution, i.e., 
\begin{equation}
P(k)_{\rm BA} = \frac{k^{-\gamma}}{\zeta(\gamma)}.
\label{eqn:bamarginal}
\end{equation}
where $\gamma$ is a constant known as the power-law exponent and $\zeta(\gamma)$ is the Riemann Zeta function of $\gamma$ which can be given as $\zeta(\gamma)=\sum_k k^{-\gamma}$.

Here we use the expression derived in \cite{Clauset2009} to estimate the power-law exponent $\gamma$ as
\begin{equation}
\gamma = 1+\tilde{N} \left[ \sum\limits_{i}^{\tilde{N}} \ln \frac{k_i}{k_{\rm min}-0.5}\right]^{-1}, 
\end{equation}
where $\tilde{N}=NP$ is the number of coupling links and $k_{\rm min}$ is the minimum degree of the coupling nodes. 
In addition, we test the goodness-of-fit under different coupling probability $P$ using the Kolmogorov-Smirnov Test, and the p-values are all larger than 0.05, implying that the fitting functions fit the data well.

\begin{figure}[t]
	\centering
	\includegraphics[width=0.24\textwidth,height=1.6in, trim=0 0 0 30]{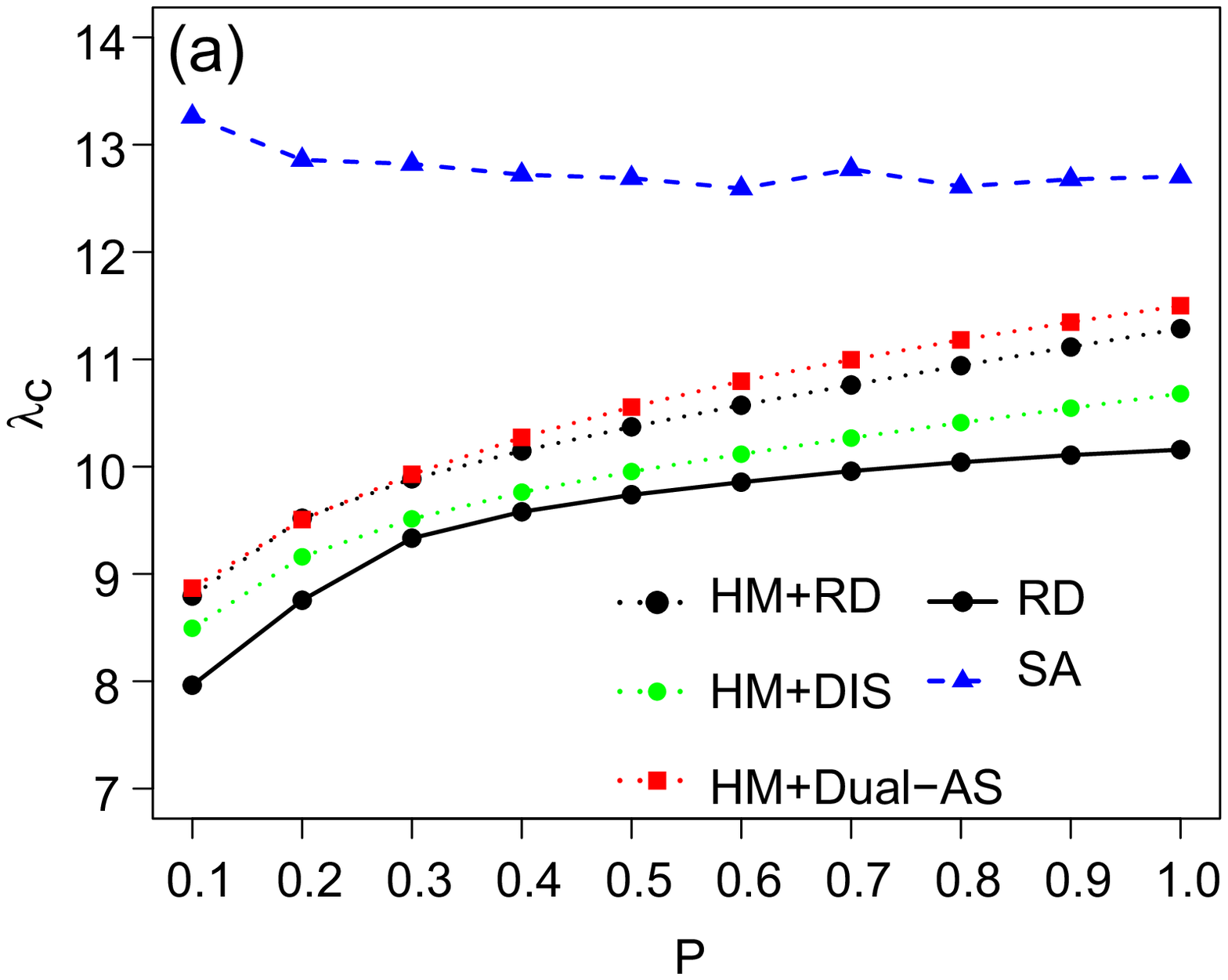}
	\vspace{-0.3cm}
	\includegraphics[width=0.24\textwidth,height=1.6in, trim=0 0 0 30]{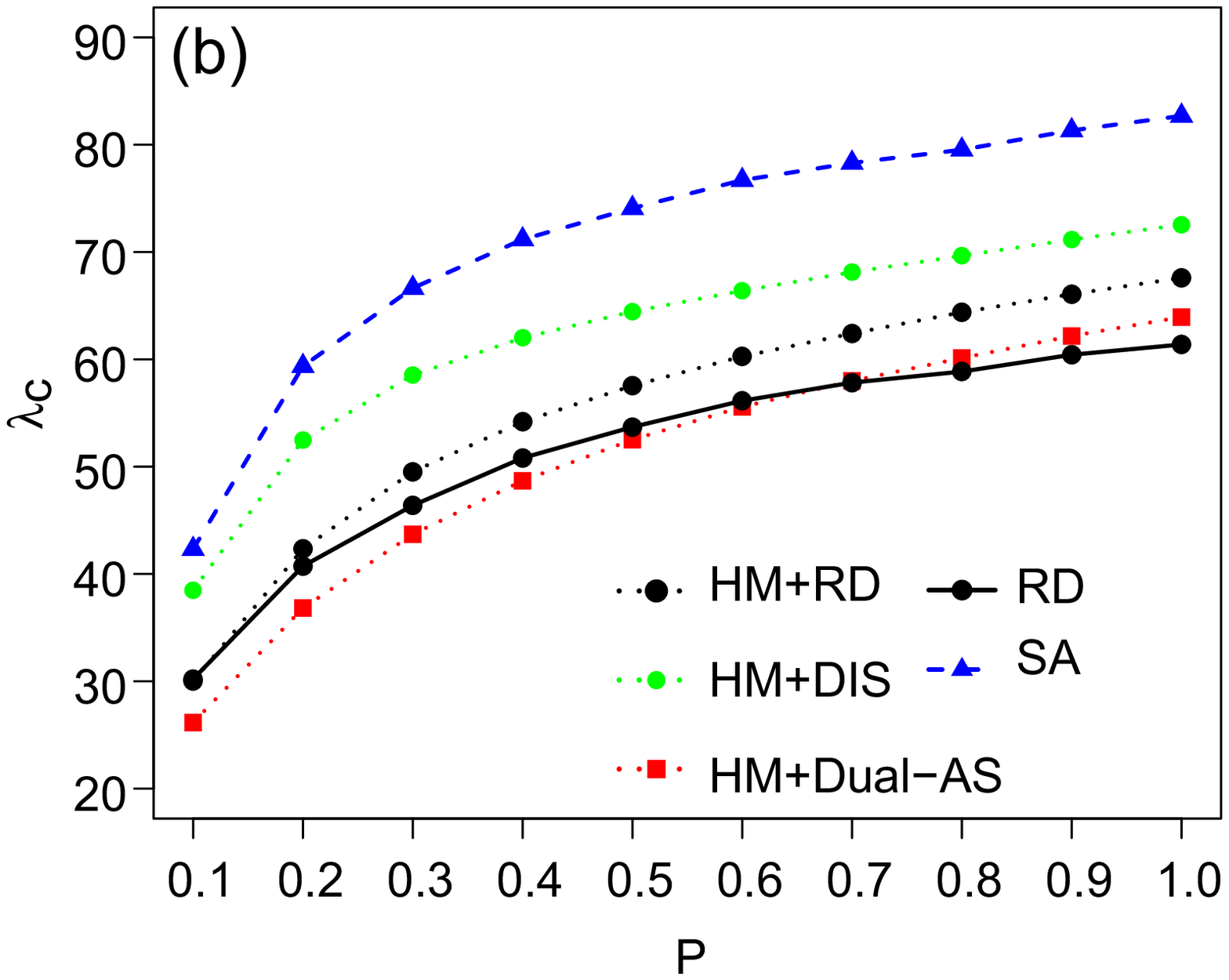}
	\caption{Critical generation rate $\lambda_c$ versus coupling probability $P$ for BA-BA(a) and ER-ER(b) interconnected networks with network size $N_A$ = $N_B$ = 500. Each point is averaged over 50 independent runs. The heuristic method outperforms the random coupling but falls behind the SA optimization algorithm in terms of $\lambda_c$.}
	\label{Fig.8}
\end{figure}

As shown in Fig.~\ref{Fig.6}(b), the degree distribution of the coupling nodes in an ER random graph looks like a Poisson distribution. Here we adopt the Nonlinear Least Square Regression method to obtain the parameter $\lambda$ of the Poisson distribution for best-fitting. Our simulation results show that under different coupling probabilities $P$, the value of $\lambda$ is always approximately equal to the average node degree of the ER random graph with very low residual standard errors. Thus the degree distribution of coupling nodes in ER random graphs can be approximated as:
\begin{equation}
P(k)_{\rm ER}  =  \frac{{\langle k \rangle}_{\rm ER}^{k}}{k!}e^{-{\langle k \rangle}_{\rm ER}}
\label{eqn:ermarginal}
\end{equation}
where ${\langle k \rangle}_{\rm ER}$ is the average node degree of the ER random graph.

Based on the above-mentioned statistical features in terms of node degree, we design the following heuristic method (HM) to add coupling links between two interconnected networks:

\begin{enumerate}
	\item Sort nodes from networks $A$ and $B$, both in ascending order of node degree. Then, we record all degree values for both networks, i.e., 
	$\left\{ D_1,D_2,...,D_{t_A}  \right\}$ and $\left\{ D_1,D_2,...,D_{t_B}  \right\}$ 
	($t_A \leq N_A$, $t_B \leq N_B$, and the degrees are not necessarily continuous) for networks $A$ and $B$, respectively.
	\item Calculate the probability of coupling selections for each degree in BA scale-free and ER random networks using Eqs.~(\ref{eqn:bamarginal}) and (\ref{eqn:ermarginal}), respectively, and then obtain the cumulative distribution probabilities $CDP(k)$ of degree $k$, for both networks A and B, i.e., 
	\begin{align*}
	&CDP(D_1)=P(D_1),\\
	&CDP(D_i)=CDP(D_{i-1})+P(D_i) 
	\end{align*}
	where $2\leq i\leq t_A$ for network $A$, and $2\leq i\leq t_B$ for network $B$.
	\item Generate a random number $r \in (0,CDP(D_{t}))$, where $D_{t}$ is the maximum degree for a network. If $r < CDP(D_1)$, randomly select a node with degree $D_1$; if $CDP(D_{i-1}) < r < CDP(D_i)$, randomly select a node with degree $D_i$. Repeat this process for network $A$ or $B$ until $NP$ nodes are determined in each side.
	\item Connect the coupling nodes from both sides under a particular coupling pattern. 
\end{enumerate}

As one can see from Fig.~\ref{Fig.6}(a), for the case of interconnected BA scale-free networks, majority of the coupling links determined by the SA algorithm are distributed in the upper-left part of the plot, while hardly any links are located in the lower-right part of the plot. Based on this observation, we design the dual-assortative (Dual-AS) coupling pattern as follows to add coupling links in step 4. In the dual-assortative (Dual-DIS) coupling, we sort the selected nodes for coupling from each network in ascending order of node degree. The other procedures of this coupling pattern is the same as the assortative coupling introduced in Section~\ref{sec:prefer}.
Additionally, we consider the random and disassortative coupling patterns in step 4 of the HM algorithm for comparison. 
Therefore, as shown in Fig.~\ref{Fig.8}, by using different coupling patterns, we design three kinds of methods, namely, HM+RD, HM+DIS and HM+Dual-AS, to add the coupling links and compare them with the RD and SA algorithms discussed in Sec.~\ref{sec:prefer}.

We can learn from Fig.~\ref{Fig.8} that the heuristic methods outperforms the random coupling but fall behind the SA optimization algorithm in terms of $\lambda_c$. As shown in Fig.~\ref{Fig.8}(a), for interconnected BA scale-free networks, the heuristic method performs best when the coupling nodes are connected using the Dual-AS coupling pattern. As for the interconnected ER random graphs, it would be better to connect the coupling nodes in a disassortative way. Moreover, with an increased coupling probability $P$, the heuristic method performs better because of a larger sample size and greater statistical significance.

\section{Conclusion\label{conclusion}}
In this paper, we considered the problem of mitigating traffic congestion in interconnected complex networks. In such networks, the way in which the coupling links are connected between the networks have significant influence on the traffic dynamics of the interconnected system. We employed a simulated annealing algorithm to find the optimal configuration of coupling links to mitigate traffic congestion. Our simulation results showed that the proposed algorithm can effectively enhance the network transmission capacity and make the networks more closely connected for both artificial interconnected networks and real-world interconnected Internet AS-level graphs, compared with assortative, disassortative and random couplings. 

Additionally, we explored the statistical features of the optimal configuration given by the SA and put forward a faster heuristic method to determine the coupling links. As a preliminary attempt, here we only considered the degree-related characteristics of the optimal patterns. 

\balance
\bibliography{bib}
\bibliographystyle{ieeetr}

\end{document}